\RequirePackage[2020-02-02]{latexrelease}
\documentclass[aps,preprint,11pt]{revtex4}
\usepackage{amsmath}
\usepackage{graphicx}
\usepackage{amsfonts}
\usepackage{amssymb}
\usepackage{blindtext}
\usepackage{multirow}
\usepackage{booktabs}
\usepackage[T1]{fontenc}
\usepackage[utf8]{inputenc}
\usepackage{float}
\usepackage{xcolor}
\usepackage{natbib}
\usepackage{subfig}
\usepackage{notoccite}
\usepackage{subcaption}
\usepackage{caption}

\RequirePackage[2020-02-02]{latexrelease}
\RequirePackage[2020-02-02]{latexrelease}
\providecommand{\U}[1]{\protect\rule{.1in}{.1in}}
\providecommand{\U}[1]{\protect\rule{.1in}{.1in}}
\providecommand{\U}[1]{\protect\rule{.1in}{.1in}}
\providecommand{\U}[1]{\protect\rule{.1in}{.1in}}
\providecommand{\U}[1]{\protect\rule{.1in}{.1in}}

\makeatletter
\renewcommand{\citation}[1]{%
	\g@addto@macro{\citation@list}{,#1}%
}
\newcommand*{\citation@list}{} % initialize
\newcommand{\sortbibitem}[2]{%
	\global\@namedef{bibitem@#1}{%
		\bibitem{#1} #2
	}%
}
\newcommand{\sort@bibitems}{%
	\@for\next:=\citation@list\do{%
		\@nameuse{bibitem@\next}%
		\global\@namedef{bibitem@\next}{}%
	}%
}
\expandafter\def\expandafter\endthebibliography\expandafter{%
	\expandafter\sort@bibitems\endthebibliography
}
\makeatother

\begin{document}
		\title{}% Force line breaks with \\
	\author{Abdelhakim Benkrane}
	\email{hakim9502.benkrane@gmail.com, abdelhakim.benkrane@univ-ouargla.dz
	}
	\address{Laboratoire LRPPS, Faculté des Mathématiques et des Sciences de la Matière.  Université Kasdi Merbah Ouargla, Ouargla 30000, Algeria}
		\date{\today}% It is always \today, today,
	\title{Entropy Analysis of Dark Matter Halo Structures}
	\begin{abstract}
In this study, we aim to derive the entropy associated with a dark matter halo modeled using a double (broken) power-law density profile. Our approach is inspired by the pioneering work of Verlinde, who proposed that gravity may not be a fundamental force but rather an emergent phenomenon rooted in entropic principles. To investigate this idea, we examine four different dark matter halo profiles: the Dehnen-type, Hernquist, Jaffe, Plummer sphere, and the perfect sphere. Using these profiles, we derive corresponding entropy expressions and assess their consistency with the second law of thermodynamics, which governs the behavior of entropy in closed systems. Furthermore, to broaden the scope of our analysis, we extend the study to a cosmological context, allowing us to explore how the derived entropy expressions influence the dynamics of the Friedmann equations that describe the evolution of the universe.

		
%In this study, we attempted to derive the entropy for a dark matter halo modeled using a double (broken) power-law density distribution. This was achieved by leveraging the seminal work of Verlinde, which is based on the idea that gravity may be an emergent force of entropic origin. We examine four specific cases, which are the Dehnen-type, Hernquist, Jaffe, Plummer sphere, and perfect sphere profiles. Then, we test the newly derived entropy expressions to check whether they satisfy the second law of thermodynamics.To broaden our analysis, we extend our study to the cosmological framework to examine how the Friedmann equations would behave with the new  form of  entropy expressions.
	\end{abstract}
	\maketitle
	\section{Introduction}
	The nature of dark matter remains one of the most profound mysteries in modern physics. Initially proposed to explain discrepancies in galactic rotation curves \cite{zwicky1933rotverschiebung, rubin1970rotation}, dark matter is now considered a fundamental component of the cosmos, accounting for approximately 27\% of the universe’s total mass-energy content \cite{yang2024black}.Despite the elusive nature of dark matter, it is very likely to play a significant and pivotal role in our understanding of cosmology and particle physics, thereby enhancing our comprehension of the universe’s structure and evolution. One of the most compelling pieces of evidence for dark matter arises from galactic dynamics. Observations of spiral galaxies reveal that their outer regions rotate at nearly constant velocities, contradicting predictions based on visible mass alone \cite{rubin1980rotational}. This anomaly suggests the presence of an unseen mass component exerting additional gravitational influence. Further evidence comes from large-scale structure formation, where numerical simulations of galaxy clustering align with a universe dominated by cold dark matter \cite{springel2005simulations}.Gravitational lensing provides another independent confirmation of dark matter’s existence. The bending of light by massive galaxy clusters, as observed in systems such as the Bullet Cluster, cannot be explained solely by visible matter \cite{clowe2006direct}. These observations indicate that a substantial, non-luminous mass component contributes to the total gravitational potential. Additionally, cosmic microwave background (CMB) measurements from experiments like WMAP and Planck further constrain dark matter’s properties, reinforcing its necessity in the $\Lambda$CDM (Lambda Cold Dark Matter) model \cite{spergel2003first}.
	Efforts to detect dark matter particles directly or indirectly remain ongoing. Theories suggest that weakly interacting massive particles (WIMPs) or axions could constitute dark matter \cite{bertone2005particle, marsh2016axion}. Underground detectors such as LUX-ZEPLIN and XENON1T aim to identify rare interactions between dark matter and ordinary matter \cite{aprile2018dark}. Complementary searches through indirect detection, including gamma-ray and cosmic-ray studies, seek to uncover annihilation signatures of dark matter particles \cite{ackermann2015searching}. \par 
	A diverse array of dark matter (DM) halo models has been proposed, informed by both simulations and astrophysical observations. Various black hole (BH) solutions have also been examined, including those incorporating a DM profile associated with a phantom scalar field \cite{Li2012}. Extensive studies have shed light on DM distribution properties \cite{Hendi2020, Rizwan2019, Shaymatov2021, Rayimbaev2021, Shaymatov2021b}, while analytical models describe supermassive BHs within DM halos \cite{Cardoso2022, Hou2018, Shen2024}. Prominent DM halo models include the Navarro-Frenk-White \cite{blas1996jf}, Einasto \cite{Dutton2014, Merritt2006}, Burkert \cite{Burkert1995}, and Dehnen \cite{Dehnen1993} models.
	
	Recently, interest has grown in studying the influence of Dehnen-type DM halos on BHs, explored from different perspectives. For example, Ref. \cite{Shukirgaliyev2021} analyzes how variations in the density profile slope affect the survival of star clusters with low star formation efficiency following rapid gas expulsion. Additionally, Ref. \cite{AlBadawi2025} investigates the impact of core density and core radius in a Dehnen-type DM halo on the environment surrounding Schwarzschild-like BHs.\par 
	On the other hand, the connection between gravity and thermodynamics is well-established, tracing back to Bekenstein, Hawking, and Jacobson's works on black holes  \cite{Bekenstein1973, Hawking1975, Jacobson1995}. This has led to the idea of gravity as an emergent phenomenon \cite{Padmanabhan2005, Padmanabhan2011, Elizalde2008, Chirco2010}, a concept further explored by Verlinde, who suggests gravity arises from an entropic force linked to changes in information about material bodies \cite{Verlinde2011}. By connecting entropy with the Bekenstein-Hawking area law, Newtonian gravity naturally emerges in this framework. While it's premature to claim gravity as an emergent phenomenon, this idea has sparked significant research into its potential implications. The general concept here is that, rather than relying on the Bekenstein-Hawking area law, one could begin with an extension of it, leading to modified forms of Newtonian gravity \cite{Sheykhi2010, Meissner2004, Sheykhi2011, Das2008, Nicolini2010, MartinezMerino2017, Beck2003, Obregon2010, Moradpour2018, Gao2010, Pazy2013, Abreu2013}
	 (see the Ref. \cite{ourabah2023other} for more details). \par 
	 In this article, we will take an approach opposite to that of the papers \cite{Sheykhi2010, Meissner2004, Sheykhi2011, Das2008, Nicolini2010, MartinezMerino2017, Beck2003, Obregon2010, Moradpour2018, Gao2010, Pazy2013, Abreu2013} and similar to Ref. \cite{ourabah2023other}; thus, we consider \cite{ourabah2023other}; thus, we will consider the gravitational potentials corresponding to the Double Power-Law Density Distributions, which describe dark matter halos and elliptical galaxies \cite{mo2010galaxy}, and finding the corresponding entropy using the Verlinde's  approach. \par 
	 This work is organized as follows:  In section \ref{entropies}, we will establish a general relationship between entropy and gravitational potential, and consequently a general relationship for the entropy of dark matter halos described by double power-law density distributions. We will then derive specific expressions that are commonly used in the literature. Then, in Section \ref{2ndlaw},
	we will verify whether the obtained entropy satisfies the second law of thermodynamics. In section \ref{fried}, our study expands by examining the impact of the derived entropy expressions on the Friedman equations. Finally, in the section \ref{conclusion}, we will present the key results obtained and discuss future prospects.
	\section{Entropies for Profile Densities of DM halo}
	\label{entropies}
	Dark matter halos are nonlinear cosmological structures with masses described by a radius-dependent density profile. Recent analytical models and simulations have improved our understanding of their properties and behavior. Numerous studies aim to derive a practical density profile that fits observational data. These efforts contribute to refining our knowledge of dark matter halos \cite{fakhry2021primordial}. 
Dark matter halos and elliptical galaxies are commonly modeled as spherical objects with a broken (double) power-law density profile, as described by \cite{mo2010galaxy},
\begin{gather}
\rho(r)=\rho_{0} \left(\dfrac{r}{r_{0}}\right)^{-\gamma}\left[1+\left(\dfrac{r}{r_{0}}\right)^{\alpha}\right]^{(\gamma-\beta)/\alpha} \label{density}, 
\end{gather}
Here, $\rho_{0}$ and $r_{0}$ represent the central halo density and radius, respectively. The parameters $(\alpha, \beta, \gamma)$ determine the type of profile as we will see later.  At small radii, the density $\rho(r)$ follows a power law with $\rho \propto r^{-\gamma}$. However, at larger radii, the density behaves as $\rho \propto r^{-\beta}$, and the parameter $\alpha$ governs the sharpness of the transition between these two regions. The model (\ref{density})  is highly flexible and encompasses various popular models as special cases, as we will see later. The total mass associated with a density distribution of the form given in Eq. (\ref{density}) is expressed as \cite{mo2010galaxy}:
\begin{gather}
		M=\dfrac{4\pi}{\alpha}\rho_{0} r_{0}^{3}\text{B}\left(\dfrac{3-\gamma}{\alpha}, \dfrac{\beta-3}{\alpha}, 1\right), \label{mass}
\end{gather}
$B(a,b,x)$ here is incomplete Beta function, which can be defined as 
\begin{gather}
	B(a,b,x)=\int_{0}^{x} t^{a-1}(1-t)^{b-1} dt. \label{key}
\end{gather}
The total mass of the system (\ref{mass}) is finite if the integral (\ref{key}) is convergent with \( x = 1 \), requiring \( a > 1 \) and \( b > 1 \). While the first condition holds for any \( \alpha > 0 \), the second requires \( \beta > 3 \). However, this constraint is not too restrictive, as several interesting models still satisfy these conditions. Examples include Dehnen’s model \( (1,4,0) \), the Jaffe profile  \( (1,4,2) \), Perfect sphere \( (2,4,0) \),   Plummer sphere \( (2,4,0) \) \cite{batic2022possible}. The corresponding gravitational potential to the double (broken) power-law density distribution is written as  
\begin{gather}
		\Phi(r)=-\dfrac{4\pi G}{\alpha}\rho_{0} r_{0}^{2}\left[\dfrac{r_{0}}{r}\text{B}\left(\dfrac{3-\gamma}{\alpha}, \dfrac{\beta-3}{\alpha}, \xi(r)\right)+\text{B}\left(\dfrac{\beta-2}{\alpha}, \dfrac{2-\gamma}{\alpha}, \xi(r)\right)\right] \label{phi},
\end{gather}
with $\xi(r)=\dfrac{(r/r_{0})^{\alpha}}{1+(r/r_{0})^{\alpha}},$  
$G$ is the gravitational constant.
In table \ref{Table1}, we present several special cases that are frequently referenced in the literature \cite{mo2010galaxy}. \par 
\begin{table}[H]
	\begin{center}
		\caption{  Double power-law density profiles.}
		\label{demo-table}
		\begin{tabular}{||c c c||} 
			\hline
			$(\alpha, \beta, \gamma)$ & Name & References \\ [0.5ex] 
			\hline\hline
			(1,3,1 ) & NFW profile &   Navarroetal.(1997) \\ 
			\hline
				(1,4,$\gamma$ ) &  Dehnen profile   &    Dehnen(1993)\\
			\hline
			(1,4,1 ) &   Hernquist profile   &     Hernquist (1990)   \\
			\hline
			(1,4,2 ) &   Jaffe profile   &      Jaffe(1983)   \\
			\hline
			(2,2,0 ) &   Modified isothermal profile   &       Sackett and Sparke(1990) \\
			\hline
				(2,3,0 ) &   Modified Hubble profile   &        Binney and Tremaine(1987) \\
				\hline
					(2,4,0 ) &  Perfect  sphere &    deZeeuw(1985) \\
					\hline
						(2,5,0 ) &  Plummer  sphere &      Plummer(1911)\\
				[1ex] 
			\hline
		\end{tabular}
	\label{Table1}
	\end{center}
\end{table}
Having outlined the gravitational potential of interest, we now turn our attention to formulating the corresponding entropic structure.\par 
Verlinde's argument \cite{Verlinde2011} considers a spherical screen of radius \( R \) centered around a massive source \( M \), with a particle of mass \( m \) placed just outside the screen. In the entropic gravity framework, the test particle interacts thermodynamically with the screen, which stores information about the massive source. As the particle moves, its displacement causes a change in entropy, leading to a restoring force, as dictated by the first law of thermodynamics,
\begin{gather}
	F\Delta x=T\Delta S,
\end{gather}
\( \Delta x \) represents the displacement of the particle from the holographic screen, while \( T \) and \( \Delta S \) denote the temperature and the entropy change on the screen, respectively. To determine the precise form of the force, one must first examine how entropy varies as the particle moves within the system. This is where the specific formulation of entropy becomes essential. By associating entropy with the area law, expressed as \( S = A / 4 \), where \( A = 4\pi r^2 \) represents the horizon area, Newtonian gravity naturally emerges \cite{Verlinde2011}. \par 
Instead of adhering strictly to the area law, we adopt a more general form of entropy, \( S(A) \), which is an arbitrary function of the surface area \( A \). This function may incorporate universal parameters, such as characteristic length scales, and in a specific limit, it reduces to the standard area law \cite{ourabah2023other}.  

The rest of the argument follows a conventional approach. We analyze a test mass \( m \) in the gravitational field of a source mass \( M \), located at the center of a spherically symmetric surface \( S \). The test mass \( m \) is assumed to be much closer to the surface than its reduced Compton wavelength, \( \lambda_m \equiv \hbar / (mc) \). When the test mass is displaced by \( \Delta x = \eta \lambda_m \) from the surface \( S \), the entropy of the surface changes by a fundamental unit \( \Delta S \), determined by the discrete spectrum of the surface's area.
\begin{gather}
	\Delta S=\dfrac{\partial S}{\partial A}\Delta A.
\end{gather}
Following the steps in  Refs. \cite{Verlinde2011,sheykhi2010entropic, ourabah2023other, sheykhi2011power}, one gets the following relation between the gravitational force   and entropy  (we impose $G=1$):
\begin{gather}
	F(r)=-\dfrac{Mm}{r^{2}} \dfrac{1}{2\pi r}\dfrac{dS}{dr} ,
\end{gather}
which can be inverted to express the entropy \( S \) in terms of the force \( F(r) \), specifically:
\begin{gather}
S=-\dfrac{2\pi}{ Mm}\int F(r) r^{3} dr.
\end{gather}
On the other hand, the gravitational force and potential are related as follows 
\begin{gather}
F(r)=-m\dfrac{d\Phi(r)}{dr},
\end{gather}
it is a straightforward exercise to show the following relation between entropy and gravitational potential 
\begin{gather}
S=\dfrac{2\pi}{M}\int r^{3}\dfrac{d\Phi(r)}{dr}dr, \label{potential}
\end{gather}
using  Eq. (\ref{phi}), one gets the expression of entropy for DM halos
\begin{gather}
S=\dfrac{2\pi  }{\text{B}\left(\dfrac{3-\gamma}{\alpha}, \dfrac{\beta-3}{\alpha}, 1\right)}r_{0}^{2}\int x^{3} \left[-\dfrac{1}{x^2} \text{B}\left(\dfrac{3-\gamma}{\alpha}, \dfrac{\beta-3}{\alpha}, \tilde{\xi}(x)\right) \right.\nonumber\\ \left.+\dfrac{1}{x}\dfrac{d \tilde{\xi}(x)}{dx}(1-\tilde{\xi}(x))^{(3-\gamma)/\alpha-1}(\tilde{\xi}(x))^{(\beta-3)/\alpha-1}
+\dfrac{d \tilde{\xi}(x)}{dx}(1-\tilde{\xi}(x))^{(\beta-2)/\alpha-1}(\tilde{\xi}(x))^{(2-\gamma)/\alpha-1}\right] dx, \label{entropy}
\end{gather}
where $x= r/r_{0}$ and $\tilde{\xi}(x)=\xi(r)=\dfrac{x^{\alpha}}{1+x^{\alpha}}.$The expression (\ref{entropy}) represents the general form of entropy associated with the gravitational potential resulting from a dark matter halo with a double power-law density profile (\ref{density}). For profiles with $\beta=3$, such as the NFW and Modified Hubble sphere, the term $\text{B}\left(\dfrac{3-\gamma}{\alpha}, \dfrac{\beta-3}{\alpha}, 1\right)$ diverges. Consequently, we will exclude this case in the next subsection, where we will compute the entropy (\ref{entropy}) for specific models.
\subsection{Dehden-type profile}
 As a first special case, let us take Dehden type profile, where $(\alpha, \beta, \gamma)=(1,4, \gamma)$. The Dehnen models are particularly useful because many of their properties can be calculated analytically, making them a powerful tool in the study of celestial objects. These properties include the intrinsic velocity dispersion for all real values of $\gamma$ between 0 and 3, and for $\gamma = 0, 1, 2$, the projected mass density and velocity dispersion as well \cite{dehnen1993family, tremaine1994family}. This model significantly contributes to understanding the structure and kinematics of galaxies. Furthermore, the Dehnen model with $\gamma = \frac{3}{2}$ closely resembles the de Vaucouleurs $r^{1/4}$ profile, commonly used to fit the surface brightness profiles of elliptical galaxies, enhancing its accuracy in simulating different galaxies and their stellar distributions. While  in  Ref. \cite{pantig2022dehnen}, a Dehnen-type dark matter halo was employed to enhance the constraints on primordial black holes as dark matter, based on ultra-faint dwarf galaxies. The corresponding entropy is written as
\begin{gather}
S(r)=2\pi r_{0}^{2}\int \dfrac{x^{3}}{2-\gamma}\frac{d\left[1-\left(\dfrac{x}{1+x}\right)^{2-\gamma}\right]}{dx}dx, \hspace{0.2cm} \text{for} \hspace{0.2cm} \gamma\neq 2,\\
S(r)=2\pi r_{0}^{2}\int x^{3} \dfrac{d \ln\left(\dfrac{x}{1+x}\right)}{dx}dx. \hspace{0.2cm} \text{for}\hspace{0.2cm} \gamma= 2
\end{gather}
By doing the two integrations, one can easily show that
\begin{gather}
S(r)=\dfrac{2\pi }{5-\gamma} r_{0}^{2}\left[\left(\dfrac{r}{r_{0}}\right)^{5 - \gamma} {}_2F_1\left(3- \gamma, 5- \gamma; 6 - \gamma; -\dfrac{r}{r_{0}}\right)\right], \label{g} \hspace{0.2cm} \text{for} \hspace{0.2cm} \gamma\neq 2, \\
S(r)=\pi  r_{0}^{2}\left[\left(\dfrac{r}{r_{0}}\right)^{2}-\dfrac{2r}{r_{0}}+2\ln\left(1+\dfrac{r}{r_{0}}\right)\right], \hspace{0.2cm} \text{for} \hspace{0.2cm} \gamma=2 \label{gg}.
\end{gather}
here, $2F_1\left(a, b; c; d\right)$ is the hypergeometric function. In terms of area $A=4\pi r^{2}$, one can write these entropies as follows
\begin{gather}
S(A)=\dfrac{2\pi }{5-\gamma} r_{0}^{2}\left[\left(\dfrac{\sqrt{A}}{\sqrt{4\pi}r_{0}}\right)^{5 - \gamma} {}_2F_1\left(3- \gamma, 5- \gamma; 6 - \gamma; -\dfrac{\sqrt{A}}{\sqrt{4\pi}r_{0}}\right)\right], \hspace{0.2cm} \text{for} \hspace{0.2cm} \gamma\neq 2, \\
S(A)=\pi  r_{0}^{2}\left[\dfrac{A}{4\pi r_{0}^{2}}-\dfrac{2\sqrt{A}}{\sqrt{4\pi}r_{0}}+2\ln\left(1+\dfrac{\sqrt{A}}{\sqrt{4\pi}r_{0}}\right)\right], \hspace{0.2cm} \text{for} \hspace{0.2cm} \gamma=2.
\end{gather}
In the case of $\gamma=2$, it can be observed that the first term is the usual form of standard entropy, $S=A/4$, while the additional terms represent corrections due to DM halo impact.
\subsection{Hernquist profile}
This profile is defined by $\alpha, \beta, \gamma=(1,4,1)$. It is a special case of Dehden-type profile with $\gamma=1$.  It describes a mass distribution that becomes steeper toward the center and declines at large radii. In addition, it is often preferred in situations where a smoother central region is required, with a more gradual falloff in density. This model  is very useful   to investigate the dynamical structure
of galaxies \cite{baes2002hernquist}. By setting \(\gamma = 1\) in relation (\ref{g}), we obtain the following entropy expression corresponding to the Hernquist profile.
\begin{gather}
S(r)=\pi \left( \frac{r (r^2 - 3r r_0 -6r_0^2)}{r + r_0} + 6 r_0^2 \ln\left(1 + \frac{r}{r_0}\right) \right),
\end{gather}
which is written in terms of event horizon area as follows
\begin{gather}
S(A)=\pi \left( \frac{\sqrt{A} \left(-\frac{A}{4\pi} + 3 \sqrt{\frac{A}{4\pi}} r_0 + 6r_0^2 \right)}{\sqrt{A} +\sqrt{4\pi} r_0} - 6 r_0^2 \ln\left(1 + \frac{\sqrt{A}}{\sqrt{4\pi}r_0} \right) \right).
\end{gather}	
\subsection{Jaffe profile}
The Jaffe profile, as originally presented in \cite{jaffe1983simple}, is another commonly used dark matter halo model, frequently applied to galaxy-scale dark matter halos.
This model is defined as ($\alpha, \beta, \gamma)=(1,4,2)$. Therefore, this is  the same as Dehden with $\gamma=2$.  When the Jaffe profile is projected onto the two-dimensional sky, it results in the de Vaucouleurs profile \cite{de1948recherches}. In fact, this profile is derived as a three-dimensional mass density profile corresponding to the projected light distributions observed in elliptical galaxies and the bulges of spiral galaxies, which are empirically well-fitted by de Vaucouleurs profiles \cite{jenny2020cosmic}.
\begin{gather}
S(r)=\pi  r_{0}^{2}\left[\left(\dfrac{r}{r_{0}}\right)^{2}-\dfrac{2r}{r_{0}}+2\ln\left(1+\dfrac{r}{r_{0}}\right)\right].
\end{gather}
%\begin{gather}
%S(r)=2\pi \left(
%\frac{2 r_0^3}{r + r_0} - 9 r_0 (r + r_0) + \frac{5}{2}  (r + r_0)^2 - \frac{1}{3r_{0}} (r + r_0)^3 + 7 r_0^2 \log(r + r_0)
%\right).
%\end{gather}
Therefore
\begin{gather}
S(A)=\pi  r_{0}^{2}\left[\dfrac{A}{4\pi r_{0}^{2}}-\dfrac{2\sqrt{A}}{\sqrt{4\pi}r_{0}}+2\ln\left(1+\dfrac{\sqrt{A}}{\sqrt{4\pi}r_{0}}\right)\right].
\end{gather}
\subsection{Plummer sphere Profile}
 The Plummer sphere can be obtained by the density (\ref{density}) by putting $(\alpha, \beta, \gamma)=(2,5,0)$. H. C. Plummer \cite{plummer1911problem} proposed a straightforward model to describe the distribution of matter within globular star clusters. The Newtonian gravitational potential associated with the Plummer sphere is expressed as
 \begin{gather}
 	\Phi(r)=-\dfrac{GM}{\sqrt{r^{2}+r_{0}^{2}}},
 \end{gather} 
For \( r \gg r_{0} \), the mass density \( \rho(r) \) decreases as \( r^{-5} \), which is steeper than what is observed in modern globular cluster models. Despite this, the Plummer model remains valuable for describing stellar density in dwarf galaxies and for simulating point particles in \( N \)-body calculations \cite{tabatabaei2024mcvittie}.\par 
For Plummer sphere profile, one easily finds the associated entropy $S$ as follows
\begin{gather}
	S=\pi\left[\dfrac{3rr_{0}^{2}+r^{3}}{\sqrt{r^{2}+r_{0}^{2}}}-3r_{0}^{2}\tanh^{-1}\left(\dfrac{r}{\sqrt{r^{2}+r_{0}^{2}}}\right)\right], \label{plum}
\end{gather}

In terms of $A$, it is easy to show that
\begin{gather}
S(A)=\pi\left[\dfrac{\sqrt{A}}{4\pi}\dfrac{12\pi r_{0}^{2}+A}{\sqrt{A+4\pi r_{0}^{2}}}-3r_{0}^{2}\tanh^{-1}\left(\sqrt{\dfrac{A}{A+4\pi r_{0}^{2}}}\right)\right].
\end{gather}
one expects for $r_{0}=0$, the usual Hawking–Bekenstein entropy; $S=\pi r^{2}=A/4$, and   this is exactly what the expression  (\ref{plum}) reduces to in the case of \( r_0 = 0 \). 
\subsection{Perfect Sphere Profile}
This model is defined by \((\alpha, \beta, \gamma) = (2, 4, 0)\). It assumes that the mass distribution in a galaxy can be described using a Stäckel-type model \cite{stackel1893ueber}, allowing for more accurate predictions of stellar motion \cite{de1985elliptical}. In stellar dynamics, the Stäckel family is a key class of integrable axisymmetric and triaxial gravitational potentials. The separability of the Hamilton-Jacobi equation in ellipsoidal coordinates ensures the existence of three independent isolating integrals of motion, making all orbits regular \cite{pascale2022regular}. The authors in  Ref.  \cite{valluri2016unified} recently analyzed a representative sample of orbits from two N-body bars and found that the dominant orbit family ($\sim$60\% of bar orbits) is the box orbit family, also known as the "non-resonant x1 orbit family" in 3D potentials. This family, arising from perturbations of the linear long-axis orbit, is crucial in the study of stationary triaxial ellipsoids, where it is the dominant orbit type \cite{de1985elliptical}.\par 
Proceeding similarly, one finds the entropic form corresponding to perfect sphere profile which reads as
\begin{gather}
S(r)=-4 \left\{ \frac{11 r r_0}{2} + \frac{2 r r_0^3}{r^2 + r_0^2} + \frac{15}{2} r_0^2 \tan^{-1} \left(\frac{r_0}{r} \right) + r^2 \tan^{-1} \left(\frac{r}{-r_0 + \sqrt{r^2 + r_0^2}} \right) 
\right\} +C(r_{0}).
\end{gather}
In terms of event horizon area, the entropy becomes
\begin{gather}
S(A)=
\frac{2}{r_0} \sqrt{\frac{A}{4\pi}} \left\{   \frac{3A}{4\pi}   \tan^{-1} \left( \frac{\sqrt{A}}{\sqrt{4\pi}r_0} \right) - r_0 \left[ \sqrt{\frac{A}{4\pi}} + \frac{4A^{3/2}}{\sqrt{\pi}\left(A+ 4\pi r_0^2\right)} + 2 r_0 \tan^{-1} \left( \frac{\sqrt{A}}{ \sqrt{A+ 4\pi r_0^2}-4\pi r_{0}} \right) \right] \right\}.
\end{gather}
So far, we have not focused on integration constants due to our lack of urgent need for them. However, since the entropy obtained in the case of a  \textbf{Perfect Sphere Profile}  (when ignoring the integration constants) is  negative for all values of \( r \) and \( r_0 \), it would be useful to add an integration constant to make the obtained entropy positive, at least for some values of \( r \) and \( r_0 \).
\section{Verification of second law of thermodynammics validation}
\label{2ndlaw}
The second law of thermodynamics asserts that the entropy of an isolated system never decreases over time; it either increases or remains constant, expressed mathematically as \( dS/dt \geq 0 \). This implies that natural processes inherently progress toward greater disorder and higher entropy. In this section, we aim to examine whether the obtained entropy expressions comply with the second law of thermodynamics.
In general, it is well known that
\begin{gather}
	\dfrac{dS}{dt}=\dfrac{dS}{dA}\dfrac{dA}{dt},
\end{gather}
Since the Hawking--Bekenstein entropy, \( S = \frac{A}{4} \), satisfies the second law of thermodynamics, it follows that \( \frac{dA}{dt} \geq 0 \). Consequently, the sign of \( \frac{dS}{dt} \) coincides with that of \( \frac{dS}{dA} \).  On the other hand, using the chain rule $dS/dA=(dS/dr) (dr/dA)$, with $dr/dA=\frac{1}{\sqrt{16\pi A}}$, so the sing of $dS/dt$ is the same as the sign of $dS/dr.$
\subsection{Dehden-type profile}
By differentiating entropies (\ref{g}) and (\ref{gg}) with respect to \( r \),   one obtains
\begin{gather}
\dfrac{dS}{dr}=2\pi r_{0} \left(\frac{r}{r_0}\right)^{5-\gamma}\left[ {}_2F_1\left(5-\gamma, -\gamma+3; 6-\gamma; -\frac{r}{r_0} \right) \frac{\gamma - 2}{\left(\frac{r}{r_0}\right)}\right.\nonumber\\\left.
+  \left(\frac{r}{r_0}\right)^{5-\gamma}(3-\gamma) \, {}_2F_1\left(-\gamma+4, 6-\gamma; 7-\gamma; -\frac{r}{r_0} \right) \frac{\gamma - 2}{(6-\gamma)} \right],  \label{en} \hspace{0.2cm} \text{for} \hspace{0.2cm} \gamma\neq 2 \\
\dfrac{dS}{dr}= \dfrac{2\pi r^{2}}{r_{0}+r}, \hspace{0.2cm} \text{for} \hspace{0.2cm} \gamma = 2. \label{forr}
\end{gather}
In both cases, \( \gamma \neq 2 \) and \( \gamma = 2 \), the second law of thermodynamics is clearly satisfied because \( dS/dr \) is positive, which ensures the expected behavior. For the case \( \gamma = 2 \) which corresponds to the Jaffe profile, \( dS/dr \)  for this profile is exactly given by Eq. (\ref{forr}).
\subsection{Hernquist Profile}
By setting $\gamma=1$, in Eq. (\ref{en}), one finds 
\begin{gather}
	\dfrac{dS}{dr}=\dfrac{2\pi r^{3}}{(r+r_{0})^{2}}, \label{30}
\end{gather}
This quantity is always positive, as expected, since we have shown that the entropy for the Dehnen-type profile satisfies the second law of thermodynamics.
%\subsection{Jaffe Profile}
%The Jaffe profile is the same as Dehnen type with $\gamma=2$, therefore
%\begin{gather}
%\dfrac{dS}{dr}=\dfrac{2\pi r^{2}}{r_{0}+r}.
%\end{gather}
\subsection{Plummer Sphere Porfile}
\begin{gather}
\dfrac{dS}{dr}=\dfrac{2\pi r^{4}}{\left(r^{2}+r_{0}^{2}\right)^{3/2}}.
\end{gather}
It is very clear that $\dfrac{dS}{dr} \geq 0$, therefore its entropy also fulfills the second law of thermodynamics.
\subsection{Perfect Sphere Profile}
\begin{gather}
	\dfrac{dS}{dr}=-\frac{4r \left(-r_0 + \sqrt{r^2 + r_0^2}\right) \left[r r_0 (5r^2 + r_0^2) + 2 (r^2 + r_0^2)^2 \tan^{-1} \left(\frac{r}{-r_0 + \sqrt{r^2 + r_0^2}}\right)\right]}{(r^2 + r_0^2)^{3/2} \left(r^2 + r_0 (r_0 - \sqrt{r^2 + r_0^2})\right)}.
\end{gather}
For the second law of thermodynamics to hold for this profile, the following condition is necessary and sufficient:
\begin{gather}
r^2 + r_0 (r_0 - \sqrt{r^2 + r_0^2}) <  0,
\end{gather}
this inequality is not satisfied. Therefore, the entropy associated with the Perfect Sphere Profile does not comply with the second law. Due to this, we will refrain from applying this profile in the modified Friedmann equation in the next section.
\section{Cosmological Implications and Modified Friedmann Equations}
\label{fried}
In this section, we extend our previous discussion to a cosmological framework and derive modified Friedmann equations based on the entropic forms introduced in section \ref{entropies}. When viewed as a consequence of a generalized entropy, these alternative gravitational theories naturally influence cosmology. Their implications can be systematically explored by utilizing the corresponding entropy expressions along with the first law of thermodynamics. Our derivation follows standard methodologies (see, e.g., \cite{sheykhi2011power, sheykhi2010thermodynamics, karami2011thermodynamics}), applied specifically to the entropic forms obtained in section \ref{entropies}.  

For this analysis, we adopt natural units by setting \( \hbar = c = k_B = G=1 \). The background spacetime is assumed to be spatially homogeneous and isotropic, described by the metric  
\begin{equation}
	ds^2 = h_{\mu\nu} dx^\mu dx^\nu + R^2 \left(d\theta^2 + \sin^2\theta d\phi^2 \right),
\end{equation}

here, \( R = a(t)r \) represents the scale factor, with coordinates \( x^0 = t \), and \( x^1 = r \). The two-dimensional metric is given by  

\begin{equation}
	h_{\mu\nu} = \text{diag} \left(-1, \frac{a^2}{1 - k r^2} \right),
\end{equation}

where \( k \) denotes the spatial curvature. The values \( k = 0, 1, -1 \) correspond to flat, closed, and open universes, respectively. A straightforward calculation yields the apparent horizon radius for the Friedmann-Robertson-Walker (FRW) universe as \cite{poisson1990internal, gong2007friedmann}  

\begin{equation}
	R = \frac{1}{\sqrt{H^2 + k/a^2}},
\end{equation}

\( H = \dot{a}/a \) is the Hubble parameter.  The matter content of the FRW universe is modeled as a perfect fluid with the stress-energy tensor  

\begin{equation}
	T_{\mu\nu} = (\rho + p) u_\mu u_\nu + p g_{\mu\nu},
\end{equation}

where \( \rho \) represents the energy density and \( p \) the pressure. This formulation leads to the continuity equation  

\begin{equation}
	\dot{\rho} + 3H(\rho + p) = 0.
\end{equation}
Using the first law of thermodynamics 
\begin{gather}
dE=TdS+WdV,
\end{gather}
and by assuming $E=\rho V$, and following the steps in Ref.\cite{sheykhi2010thermodynamics, ourabah2023other}, one can easily get the relation between energy density and entropy
\begin{gather}
 \rho=-\dfrac{3}{\pi}\int \dfrac{\partial S}{\partial A}\bigg\arrowvert_{A=4\pi R^{2}}\dfrac{dR}{R^{3}}+C, \label{1st}
\end{gather}
\( C \) is an integration constant, which will later be determined by ensuring that the standard Friedmann equation is recovered in the appropriate limit, especially  in the case of Plumer sphere profile, where the limit $r_{0} \rightarrow 0$, implies  the standard Friedmann equation, since in this limit, we recover  the usual entropy expression $S=A/4=\pi r^{2}$ .
\subsection{Dehden-type profile}
Using  Eqs. (\ref{en}), (\ref{forr}) and (\ref{1st}), one can easily show that the generalized Friedmann equation for this profile is written as
\begin{gather}
\dfrac{8\pi \rho}{3}=\frac{8 \left( 1+\sqrt{H^2 + \frac{k}{a^2}}  r_0 \right)^{\gamma - 2} \left(r_{0}(\gamma-2)\sqrt{H^2 + \frac{k}{a^2}}-1\right) }{r_0^2 (\gamma - 2)(\gamma - 1)}+C(\gamma, r_{0}), \hspace{0.2cm} \text{for}\hspace{0.2cm} \gamma\neq 2,  \label{expression}\\
\dfrac{8\pi \rho}{3}=\frac{2 \left( r_0 \sqrt{H^2 + \frac{k}{a^2}} - \ln\left( 1 + r_0 \sqrt{H^2 + \frac{k}{a^2}} \right) \right)}{r_0^2}, \hspace{0.5cm} \text{for} \hspace{0.2cm} \gamma= 2,
\end{gather}  
we will discuss the constant  
\( C(\gamma, r_{0}) \)  
in more detail later when we address a special case of the Dehnen type, specifically the case  
\( \gamma = 1 \)   (Hernquist Profile). It is important to note that the modified Friedmann equation in the Dehnen case for $\gamma=2$ is the same to the one in the Jaffe profile, as already mentioned. Therefore, we will not address the modified Friedmann equation for the Jaffe profile separately.
\subsection{Hernquist Profile}
Using the Eqs. (\ref{30}) and (\ref{1st}), one obtains   the generalized Friedmann equation for Hernquist profile 
\begin{gather}
\dfrac{8\pi \rho}{3}=
\frac{2 \left( \ln \left( 1 + r_0 \sqrt{H^2 + \frac{k}{a^2}} \right) - \frac{\sqrt{H^2 + \frac{k}{a^2}}r_0}{1 +\sqrt{H^2 + \frac{k}{a^2}} r_0} \right)}{r_0^2}. \label{hern}
\end{gather}
As we mentioned earlier, the Hernquist profile is a special case of Dehnen profile, with $\gamma=1$, based on this, one can put constraint on the constant $C(\gamma, r_{0})$ in such  Eqs (\ref{expression}) and (\ref{hern}) coincide at $\gamma=1$. Therefore 
\begin{gather}
\lim_{\gamma \rightarrow 1}\left\{\frac{8 \left( 1+\sqrt{H^2 + \frac{k}{a^2}}  r_0 \right)^{\gamma - 2} \left(r_{0}(\gamma-2)\sqrt{H^2 + \frac{k}{a^2}}-1\right) }{r_0^2 (\gamma - 2)(\gamma - 1)}+C(\gamma, r_{0})\right\}\nonumber\\
=\frac{2 \left( \ln \left( 1 + r_0 \sqrt{H^2 + \frac{k}{a^2}} \right) - \frac{\sqrt{H^2 + \frac{k}{a^2}}r_0}{1 +\sqrt{H^2 + \frac{k}{a^2}} r_0} \right)}{r_0^2}  \normalsize.\label{constarint}
\end{gather}
The first term in Eq.~(\ref{constarint}) diverges as \( \gamma \to 1 \). To ensure that the full expression remains finite, the constant \( C(\gamma, r_0) \) must exhibit a compensating divergence with the opposite sign.
\subsection{Plummer Sphere Profile}
\begin{gather}
\dfrac{8\pi \rho}{3}=-\dfrac{2}{r_{0}^{2}\sqrt{1 + \left(H^2 + \frac{k}{a^2}\right)r_0^2}}+C(r_{0}),
\end{gather}
the constant $C(r_{0})$ must ensure that the energy density remains positive as well in the limit $r_{0}\rightarrow0$ we recover the standard form of the Friedmann equation, since at this limit, we get the usual entropy ($S=A/4=\pi r^{2}$), as we mentioned earlier about this profile. The standard Friedmann equation is 
\begin{gather}
\dfrac{8\pi \rho}{3}=H^{2}+\dfrac{k}{a^{2}}.
\end{gather}
 The only form of the constant $C(r_{0})$ that satisfies these requirements  is  $C(r_{0})=\dfrac{2}{r_{0}^{2}}.$ Therefore, the final form of the generalized Friedmann for Plummer sphere profile is written as
\begin{gather}
\dfrac{8\pi \rho}{3}=\dfrac{2}{r_{0}^{2}}\left\{-\dfrac{1}{\sqrt{1 + \left(H^2 + \frac{k}{a^2}\right)r_0^2}}+1\right\}.
\end{gather}
\newpage
\section{Conclusion}
\label{conclusion}
This study derived the entropy corresponding to a double (broken) power-law density distribution, which represents models of dark matter halos. First, we established a general relationship between entropy and the density profile parameters \((\alpha, \beta, \gamma)\) that characterize the dark matter halo. Subsequently, we applied this framework to specific models, including the Dehnen-type, Hernquist, Jaffe, Plummer sphere, and perfect sphere profiles.  \par 
Next, we examined whether the obtained entropy expressions satisfy the second law of thermodynamics. We found that the law holds in all cases except for the entropy corresponding to the perfect sphere profile.\par 
As a final step in this work, we derived the modified Friedmann equations in the presence of a dark matter halo.  One potential approach is to apply observational constraints to the modified Friedmann equations explored in this study, and assess their compatibility with current cosmological data.\par 
Our study has the potential to open significant future directions, where  the new class of entropies derived here can be used to explore their implications for black hole thermodynamics— offering  promising opportunities for  deeper understanding, as mentioned in \cite{ourabah2023other}. In addition, since this work is within the context of dark matter halos, the derived entropy expressions may provide deeper insights into their properties.
\vspace{0.5cm}

\textbf{Data Availability Statement:}: No Data associated in the manuscript\par 
The authors declare that they have no known competing financial interests or personal relationships that could
have appeared to influence the work reported in this article
\bibliographystyle{unsrt} 
\bibliography{Darkreff} 
\end{document}